\pdfoutput=1
\documentclass[12pt]{article}
\usepackage[utf8]{inputenc}
\usepackage{geometry}
\usepackage[square,numbers,sort]{natbib}
\usepackage{graphicx}
\usepackage{changepage}
\usepackage{multirow}
\usepackage{color}
\usepackage{caption}
\usepackage{subcaption}
\usepackage{tabularx}
\usepackage{array}
\usepackage{booktabs}
\usepackage{array} 
\usepackage{authblk}
\newcolumntype{P}[1]{>{\raggedright\arraybackslash}p{#1}}

\newcommand{\revision}[1]{\textcolor{black}{#1}}
\newcommand{\italquote}[1]{\begin{quote}``\textit{#1}''\end{quote}}
\newcommand{\redacted}[1]{\textbf{REDACTED}}

\title{Understanding the Role of Large Language Models in Personalizing and Scaffolding Strategies to Combat Academic Procrastination}

\author[1]{Ananya Bhattacharjee}
\author[1]{Yuchen Zeng}
\author[1]{Sarah Yi Xu}
\author[2]{Dana Kulzhabayeva}
\author[1]{Minyi Ma}
\author[3]{Rachel Kornfield}
\author[1]{Syed Ishtiaque Ahmed}
\author[1]{Alex Mariakakis}
\author[4]{Mary P Czerwinski}
\author[5]{Anastasia Kuzminykh}
\author[6]{Michael Liut}
\author[1, 2]{Joseph Jay Williams}

\affil[1]{Computer Science, University of Toronto, Toronto, Ontario, Canada}
\affil[2]{Psychology, University of Toronto, Toronto, Ontario, Canada}
\affil[3]{Preventive Medicine, Northwestern University, Chicago, Illinois, United States}
\affil[4]{Microsoft Research, Redmond, Washington, United States}
\affil[5]{Faculty of Information, University of Toronto, Toronto, Ontario, Canada}
\affil[6]{Mathematical and Computational Sciences, University of Toronto Mississauga, Mississauga, Ontario, Canada}
\date{}

\begin{document}
\maketitle
\begin{abstract}
  \noindent
Traditional interventions for academic procrastination often fail to capture the nuanced, individual-specific factors that underlie them. Large language models (LLMs) hold immense potential for addressing this gap by permitting open-ended inputs, including the ability to customize interventions to individuals' unique needs. However, user expectations and potential limitations of LLMs in this context remain underexplored. To address this, we conducted interviews and focus group discussions with 15 university students and 6 experts, during which a technology probe for generating personalized advice for managing procrastination was presented. Our results highlight the necessity for LLMs to provide structured, deadline-oriented steps and enhanced user support mechanisms. Additionally, our results surface the need for an adaptive approach to questioning based on factors like busyness. These findings offer crucial design implications for the development of LLM-based tools for managing procrastination while cautioning the use of LLMs for therapeutic guidance.
\end{abstract}



\maketitle

\noindent\textbf{Keywords:} Procrastination, Personalized Reflections, GPT-4, ChatGPT, Large Language Models, Education

\section{Introduction}

Procrastination, defined as the intentional delay of tasks despite known negative consequences \cite{pychyl2013solving, arakawa2023catalyst, day2000patterns}, is especially widespread in academic settings. Prior research indicates that at least half of all college students engage in chronic procrastination \cite{steel2007nature, grunschel2015students}, which can negatively impact their academic performance and overall wellbeing. Researchers have developed a range of interventions for procrastination management, focusing on task completion \cite{arakawa2023catalyst, inie2021aiki}, time management \cite{inie2021aiki, kim2016timeaware}, and personalized motivation \cite{rozental2015internet, toker2015effect}. However, they frequently fail to capture the nuanced factors that contribute to procrastination, such as individual learning styles and environmental influences \cite{bhattacharjee2023investigating, kovacs2019conservation}. The challenge lies in these factors being subjective and difficult to quantify, rendering traditional interventions less effective.




Large language models (LLMs) have recently garnered significant attention for their capability to act as personalized intervention tools and supportive scaffolds in numerous settings \cite{abd2023large, arakawa2023catalyst, wang2023enabling, kumar2023exploring, uchiyama2023large, liu2022design}. With their computational prowess and linguistic sophistication, these models can analyze a vast array of variables via text-based input, potentially offering a level of personalization and support that is difficult to achieve with traditional intervention methods. \revision{LLMs also stand out due to their dynamic interaction capabilities, surpassing the constraints of generic surveys and rule-based chatbots that lack conversational depth and follow-up capabilities.
In the context of academic procrastination, the potential of LLMs lies not in replacing traditional methods like setting reminders or automatic scheduling, but in their ability to contextualize and intelligently integrate a broad spectrum of procrastination management strategies.} 

By leveraging extensive datasets and natural language processing techniques, LLMs may possess the ability to craft advice and recommendations that are highly tailored to an individual's unique academic and personal circumstances \cite{lin2022applying, mogavi2023exploring}.  
For instance, an LLM could ask a student about their upcoming deadlines and work schedule to collaboratively help them generate micro-deadlines that fit into their schedule. Knowing that the student has two term papers and a part-time job, the LLM might advise dedicating the first two hours after work on a specific day to outlining one of the term papers. This sort of targeted advice has the potential to acknowledge students' unique circumstances and constraints, thereby increasing the likelihood of the advice being both practical and effective.

However, LLMs cannot be trivially adapted for such intervention personalization. 
\revision{They currently lack the innate capacity to perceive users' emotional states or to navigate the complexities and subjectivity of intricate social contexts (although this may change with future advancements) \cite{xi2023rise}. Therefore, users need to provide such nuanced information for LLMs to offer effective personalized recommendations, and the applicability of the recommendations they receive depends substantially on the depth of the information provided \cite{liu2022design}.}
It is also important for researchers to understand which intervention elements users wish to personalize. 
For instance, in the context of academic procrastination, users may desire an option to specify the type of support they need, such as reminders for deadlines versus tips for time management. 
Insights into these user expectations can guide the design of prompts to elicit targeted and useful responses through prompt engineering, aligning closely with what users seek in managing procrastination \cite{wang2023prompt, arvidsson2023prompt}. 
They may even help adapt the attention mechanisms within LLMs \cite{vaswani2017attention} to focus on certain keywords or contextual clues within a user's query or text input, such as deadlines, emotional states, or specific tasks.
Without such insights, there is a risk that developers create LLM-based tools that are misaligned with user needs and expectations \cite{bhattacharjee2023design, kabir2022ask}. Finally, an understanding of the limitations and guardrails required for LLM-based tools in this space is also important to ensure that intervention adjustments are still consistent with validated strategies or clinical guidelines.



Motivated by both the challenges and opportunities in applying LLMs to intervention personalization, as well as their potential to act as a supportive scaffold, we explore how students can use an LLM-based tool to manage academic procrastination. Specifically, our focus extends to task initiation, deadline management, emotional regulation, and addressing the impact of social and environmental factors on academic performance. We aim to understand how users envision the ways in which LLMs might facilitate these strategies for managing academic procrastination. 
At the same time, we attempt to identify potential limitations of LLMs and design tensions that should be considered for future tools in this space. Our work is motivated by the following research questions:\\\vspace{-0.5em}

\begin{adjustwidth}{4em}{}
\begin{itemize}
    \item[\textbf{\textbf{RQ1}:}] \revision{How do users envision the role of LLMs in tailoring strategies for managing academic procrastination?}
    \item[\textbf{\textbf{RQ2}:}] What challenges or tensions might arise when using LLMs for tailoring strategies to manage academic procrastination?\\\vspace{-0.5em}
\end{itemize}
\end{adjustwidth}

In this study, we interacted with 15 university students through individual interviews and focus groups to delve into how they perceive the potential of LLM-based tools in mitigating academic procrastination. Before posing questions, we offered participants an opportunity to engage with a functional technology probe \cite{hutchinson2003technology, boehner2007hci}, specifically designed to incorporate strategies for managing procrastination. This probe served as a contextual anchor for their subsequent recommendations for the design of LLM-based tools for procrastination management. To bolster our results, we also consulted six experts from the fields of clinical psychology, education, and cognitive science to validate our findings and explore potential design challenges.


\revision{Our study illuminates key preferences and concerns regarding the use of LLM-based tools in managing academic procrastination. Participants highlighted a preference for structured, deadline-focused planning coupled with real-world examples for guidance. In contrast, experts advised caution regarding the scope of these tools, emphasizing the need for emotional validation and critical thinking rather than therapeutic intervention. These findings enhance our understanding of human-LLM collaboration, advocating for a balance between methodical task management and adherence to ethical standards in emotional support. The design recommendations derived from our study call for flexible guidance systems and seamless integration into daily routines while setting clear boundaries for the emotional aspects of procrastination management. Overall, our work not only identifies specific user requirements for managing academic procrastination but also contributes to the discourse on the ethical and efficient deployment of LLM-based tools in diverse contexts.}
\section{Related Work}

We first delve into existing literature concerning the application of LLMs in academic environments, focusing specifically on their use in personalization and contextual adaptation, their role as a supportive scaffold, and the ethical challenges that arise from their use. Subsequently, we turn our attention to prior interventions aimed at mitigating academic procrastination.

\subsection{Application of LLMs in Academic Settings}

The emergence of LLMs has introduced new possibilities in learning and education. Within academic settings, they have the potential to serve multiple roles — from personalizing learning experiences to providing task-oriented support \cite{mogavi2023exploring, tian2023last, arakawa2023catalyst, yuan2022wordcraft, abd2023large}. As a result, researchers have started exploring the role of LLM-based tools in different phases of education, including content creation and summarization, student evaluation, collaborative data analysis, and research \cite{dang2023choice, mogavi2023exploring, abd2023large, wang2023next, halaweh2023chatgpt, kumar2023exploring, kumar2023math}.

One of the most salient strengths of LLMs in educational settings is their capacity for personalization and contextual adaptation \cite{wang2023enabling, kumar2023exploring, dang2023choice, kong2023investigating}. It has sparked several potential application areas in the academic arena, although much of the current research is exploratory in nature and yet to be validated on a large scale \cite{kasneci2023chatgpt}.
Educational platforms equipped with LLMs have been used to analyze the historical data of a student's essay submissions, allowing them to suggest areas of improvement \cite{uchiyama2023large} and to identify recurring grammatical errors or stylistic issues \cite{dang2023choice, moore2023empowering, mogavi2023exploring, williamson2023re}. In programming courses, LLMs might act as adaptable tutors that provide hints or suggest debugging strategies when students encounter errors in their code \cite{liffiton2023codehelp, kazemitabaar2023studying, denny2023promptly, leinonen2023using}. Unlike static FAQs or forums, the generated advice can take into account the specific syntax and logic of the student's work. Furthermore, LLMs have the potential to facilitate personalized learning journeys by recommending courses or materials that correspond with an individual's academic objectives and interests \cite{kasneci2023chatgpt, bhutoria2022personalized, morreel2023aye}. To illustrate, if a student indicates a desire to achieve proficiency in a specific programming language, an LLM could suggest a series of progressively challenging exercises and relevant readings, thereby offering a bespoke pathway toward mastery \cite{mogavi2023exploring}.

Beyond their capacity for personalization, LLMs also have the potential to excel as supportive scaffolds in academic tasks that require additional guidance or collaborative input \cite{lee2022coauthor, yuan2022wordcraft, arakawa2023catalyst, zhu2022action, kim2022learning, denny2023conversing, imani2023industry}, although the research is still premature. In writing-intensive courses, LLMs might function as intelligent writing assistants \cite{lee2022coauthor, gmeiner2023dimensions, kim2023towards, wei2022chain}; they can provide more than mere copy-editing, but also real-time support in enhancing the structure and flow of arguments. They may also assist students in brainstorming ideas, organizing their thoughts, and co-creating outlines for essays or research papers \cite{kasneci2023chatgpt, abd2023large, zhang2023benchmarking, tang2023evaluating, kitamura2023chatgpt}. Additionally, LLMs can be helpful for microlearning experiences, particularly in subjects that involve complex scientific concepts \cite{lin2022applying, mogavi2023exploring}. By breaking down complex ideas into more digestible and comprehensible explanations, they might offer students a way to incrementally strengthen their understanding.

While the promises of LLMs in academic settings are manifold, these benefits come with ethical concerns that cannot be ignored \cite{lund2023chatgpt, thirunavukarasu2023large, jakesch2023co, kasneci2023chatgpt, jo2023understanding, bender2021dangers}. One significant worry is the potentially detrimental impact LLMs can have on critical thinking and problem-solving skills \cite{kasneci2023chatgpt, jakesch2023co}. The ease and immediacy of AI-generated responses may discourage students from delving deeply into subjects, opting instead for quick answers without a solid understanding of underlying principles \cite{mogavi2023exploring, jakesch2023co, milano2023large}. This issue is exacerbated by the possibility that students may rely too heavily on LLMs, diminishing traditional student-teacher interactions that are often essential for learning \cite{milano2023large}. There is also the problem of accuracy \cite{dale2022detecting, pagnoni2021understanding, pavlik2023collaborating}. Students might not question the validity of the information provided by LLMs under the false assumption that any machine-generated information must be correct. This could lead to the unintentional dissemination of false or misleading information \cite{thieme2023foundation, korngiebel2021considering}. Such concerns could be heightened in the specific context of academic procrastination management, where erroneous or contextually inappropriate advice could have detrimental effects \cite{jo2023understanding, spatharioti2023comparing}.  
These ethical issues bring into focus the need for careful design and implementation strategies when incorporating LLMs into academic environments.




\subsection{Interventions to Manage Procrastination}

Efforts to alleviate procrastination typically aim to close the distance between intending to act and actually taking action \cite{sheeran2016intention, gollwitzer2006implementation}. These efforts are multifaceted, incorporating tactics that improve self-regulation, bolster self-efficacy, and nurture social support networks \cite{van2018overcoming, schouwenburg2004perspectives}. Various modalities have been employed to deliver these strategies, such as self-reflection applications, interactive conversational systems, technology-based interruptions, life management software, structured workshops, and therapeutic sessions \cite{pychyl2013solving, van2018overcoming, kovacs2019conservation, zavaleta2022can, arakawa2023catalyst}.

A large body of work on procrastination management has focused on enhancing self-regulatory skills \cite{posner2000developing}, through nurturing positive habits like goal setting and time management \cite{steel2007nature, cirillo2018pomodoro, kim2016timeaware}. Various tools have been designed to help individuals spend more time on focused tasks, thereby aiding them in meeting their goals and deadlines \cite{pereira2021struggling, grover2020design, brechmann2013remind, das2023focused, hsieh2008using, kovacs2019conservation}. For instance, \citet{pereira2021struggling} examined the effectiveness of GanttBot, a Telegram chatbot designed to reduce procrastination in single-student capstone projects. Incorporating various features like alerts, advice, automatic rescheduling, motivational messages, and historical project references, the chatbot was successful in reducing the number of overdue days among students compared to a control group. A chatbot developed by \citet{grover2020design} that allowed information workers to block out time in calendars and reflect on daily mood also saw similar success in helping them spend more time on focused tasks. 

Another prominent line of work has investigated the use of reflection and reminders as mechanisms to promote timely action and adherence to deadlines \cite{edwards2015examining, martin2015effects, baker2016randomized, humphrey2021exploring, ye2022behavioral, zavaleta2022can}. For instance, \citet{kim2016timeaware} introduced a system that encouraged self-reflection on periods of distraction, noting a moderate decrease in procrastinatory behavior among users. Similarly, \citet{edwards2015examining} used email reminders to inform students about their prior homework performance, which led to a reduction in late submissions. Further evidence from \citet{ye2022behavioral} and \citet{humphrey2021exploring} suggests that both email and text message reminders can offer moderate benefits in promoting timely submissions and improving academic performance.

Other strategies for combating procrastination included direct technological interventions and the leveraging of social support \cite{cirillo2018pomodoro, kim2016timeaware, arakawa2023catalyst, tuasikal2019role, hsueh2020exploring}. For example, \citet{inie2021aiki} introduced a browser extension that prompts users with microlearning activities when they attempt to visit time-wasting websites, thereby redirecting their focus. Similarly, \citet{arakawa2023catalyst} employed generative AI models to help individuals pick up where they left off on interrupted tasks by providing contextual cues, thus aiding in the transition back to the activity at hand. On the social front, studies have underscored the importance of social support in curbing procrastination \cite{uzun2013reducing, tuasikal2019role, hsueh2020exploring}. \citet{tuasikal2019role} demonstrated that a higher level of social support  could lead to less procrastination in thesis writing, while \citet{hsueh2020exploring} found that being part of a larger group helped reduce distractions and improve goal-setting.



However, a recurring critique of all of these interventions from procrastination management is their frequent inability to address the complex, individualized factors that contribute to procrastination \cite{kovacs2019conservation, bhattacharjee2023investigating}. Factors such as social circumstances, motivational triggers, or unique lifestyles are often challenging to incorporate into these designs. Addressing such nuanced issues could potentially benefit from the open-ended input capabilities of LLMs. However, there is a notable gap in understanding how users might perceive and interact with these LLM systems, how LLMs could enhance existing interventions, and what challenges and ethical considerations may be associated with their use. Our study aims to delve into these issues.

\section{Design of the Probe}
\label{ref: design}
In order to probe student participants' reactions to how LLMs can be used to help them address procrastination, we developed a web-based technology probe called the Self-generated Personalized Articulations and Reflections Kit (SPARK).
SPARK utilizes OpenAI's GPT-4 model \cite{OpenAI_2023} to offer context-specific advice for managing procrastination.
The tool was iteratively designed by our research team composed of faculty members and graduate students in human-computer interaction, natural language processing, cognitive science, and psychology. 

While the design of SPARK was informed by existing literature (as described later in this section), our primary goal was not to propose and validate its design. Instead, our primary goal was to understand the features and affordances users would find valuable in such a tool. We believed that asking participants to generate design ideas in a vacuum could be challenging \cite{liao2022connecting, bhattacharjee2023investigating, thieme2023foundation}; therefore, we provided them with SPARK as a probe to elicit more informed responses.
SPARK comprises four key components, each of which we elaborate upon below.

\subsection{Seed Message} 

The tool initially presented participants with a message grounded in the existing literature on procrastination management \cite{pychyl2013solving, scent2014acceptance, posner2000developing, bandura1994self, uzun2013reducing}. This message served as a generic blueprint for addressing procrastination and involved a subset of four strategies from prior work~\cite{pychyl2013solving, scent2014acceptance}: 
\begin{itemize}
\item \textbf{Cognitive Insight:} This component educates users about the complex psychological and cognitive factors contributing to procrastination, such as fear of failure and temporal discounting bias
\item \textbf{Psychological Flexibility:} This component encourages the adoption of a flexible mental stance, promoting acceptance and mindfulness. It suggests that individuals adopt the role of an observer of their own emotional and cognitive states.
\item \textbf{Value Alignment:} This component directs users to contemplate how completing the task correlates with their core values and highlights the potential costs of not adhering to these values.
\item \textbf{Implementation Intention:} This component advocates for the establishment of actionable plans to align behavior with one's values when certain thoughts or emotional triggers arise.
\end{itemize}

\begin{figure}[h!]
\centering
\includegraphics[width=11cm]{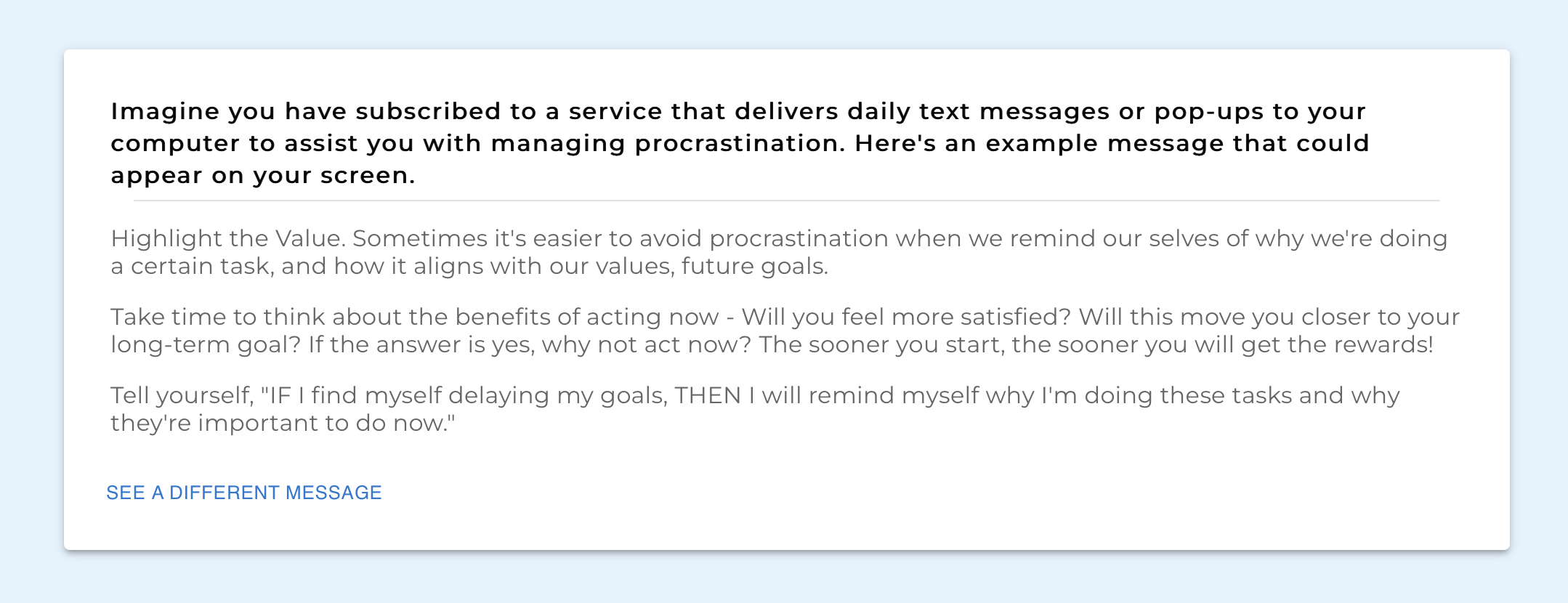}
\caption{The screen through which participants saw the seed message}
\label{fig: seed message}
\end{figure}

\noindent
Figure \ref{fig: seed message} illustrates how seed messages were presented, and Table \ref{tab: seed messages} provides some examples of the seed messages with associated psychological principles.

\begin{table}
\centering
\caption{Example Seed Messages. The psychological principles indicated by arrows were not part of the main text but are included here to annotate the different components of each message.}
\label{tab: seed messages}
\begin{tabular}{|l|p{13cm}|}
\hline
\textbf{Number} & \textbf{Complete Message (and Associated Psychological Principles)} \\
\hline
1 & Highlight the Value. Sometimes it's easier to avoid procrastination when we remind ourselves of why we're doing a certain task, and how it aligns with our values, future goals.   \textcolor{blue}{ \textrightarrow } \textbf{Cognitive Insight}

Take time to think about the benefits of acting now - Will you feel more satisfied? Will this move you closer to your long-term goal? If the answer is yes, why not act now? The sooner you start, the sooner you will get the rewards! \textcolor{blue}{ \textrightarrow } \textbf{Value Alignment } 

Tell yourself, ``IF I find myself delaying my goals, THEN I will remind myself why I'm doing these tasks and why they're important to do now.''  \textcolor{blue}{ \textrightarrow } \textbf{Implementation Intention}\\
\hline
2 &   Future Is Real. We can know something without realizing it - without it becoming ``real'' and relevant to our lives. When you procrastinate on exercising, you are likely aware of the negative consequences on your physical health, but you may not really care about it because the negative consequences can take fairly long to be detrimental. \textcolor{blue}{ \textrightarrow } \textbf{Cognitive Insight}

However, that day will come sooner or later. When you are about to procrastinate, think about the negative consequences and how they snowball. The task may not seem pressing now, but would it still be fine if you keep procrastinating on it? How would it change your life trajectory? The future is not far away, it is REAL. Don’t leave guilt and regrets until it is too late. \textcolor{blue}{ \textrightarrow } \textbf{Psychological Flexibility}

Tell yourself, ``IF I find myself putting off a task that doesn't seem pressing, THEN I will remind myself that things often don't seem pressing until it's too late. I will get started on the task.'' \textcolor{blue}{ \textrightarrow } \textbf{Implementation Intention}
 \\
\hline
3 &   Own Yourself. Overcoming procrastination is about self-regulation. We know we shouldn't procrastinate, but we are unable or unwilling to regulate ourselves to act because procrastination provides immediate mood repair.  \textcolor{blue}{ \textrightarrow } \textbf{Cognitive Insight}

When you are about to give in to feel good, always remind yourself - you are the owner of yourself, you can control yourself, and you should control yourself. You wouldn't wish to sacrifice something more important for immediate mood repair. You wouldn't let desires take control over you.  \textcolor{blue}{ \textrightarrow } \textbf{Psychological Flexibility}

Tell yourself, ``IF I find myself saying "I feel like it tomorrow", THEN I know it signals giving in, and I will just start on the task that I'm procrastinating.''\textcolor{blue}{ \textrightarrow } \textbf{Implementation Intention}
 \\
\hline
\end{tabular}
\end{table}

\subsection{Interface Options to Generate Context Specific Messages by LLMs} 
\begin{figure}[h!]
\centering
\includegraphics[width=13.5cm]{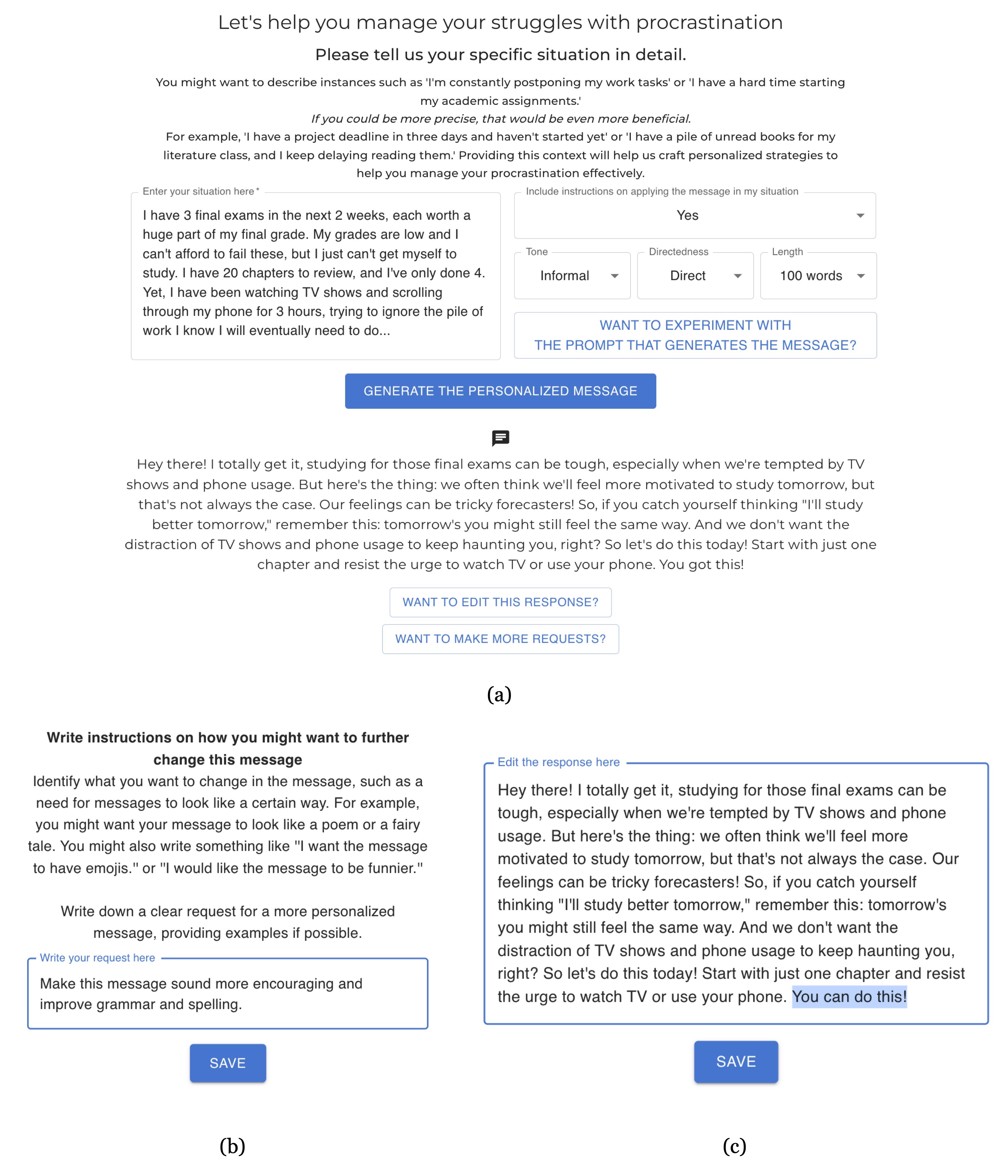}
\caption{Interactive features for generating context-specific messages: (a) customizable features that allow users to describe their individual situation, select the need for specific instructions, choose the tone and directedness of the message, and specify the desired length of the message; (b) a feature enabling open-ended modifications to generated messages; (c) a built-in editor that allows for direct edits to the generated message.}
\label{fig: buttons}
\end{figure}

To gather nuanced insights into the challenges individuals face with procrastination, we provided users with an open-ended text box where they could freely share details they considered pertinent. Users also had the option to specify the tone (formal vs. informal), directedness (direct vs. indirect), and length of the LLM-generated message (50, 100, or 150 words). They could choose whether or not to include specific instructions for addressing their situation, as well as select specific prompts for generating the message (more about this in Section \ref{instructions}). These options were informed by prior literature on personalized advice \cite{malle2001attention, bandura1994self, gnambs2010color}. Figure \ref{fig: buttons}(a) illustrates how these options were presented.

The structure for generating each tailored response was dictated by a specific query format: ``Adapt and rewrite [Seed Message] for [User Name], who is facing the following challenge: [User Situation]. The rewritten message should be [Selected Tone], [Selected Directedness], and have fewer than [Selected Length] words. [Include instructions on applying the message in the situation].''  In this query, the final sentence would be excluded if the user had no preference for receiving specific application instructions for their particular situation. This query underwent multiple iterations to achieve its final form.

Participants also had the option to make open-ended requests to change the generated message as well as edit the message themselves. Figures \ref{fig: buttons}(b) and \ref{fig: buttons}(c) show how these features were presented, respectively.


\begin{figure}[h!]
\centering
\includegraphics[width=13cm]{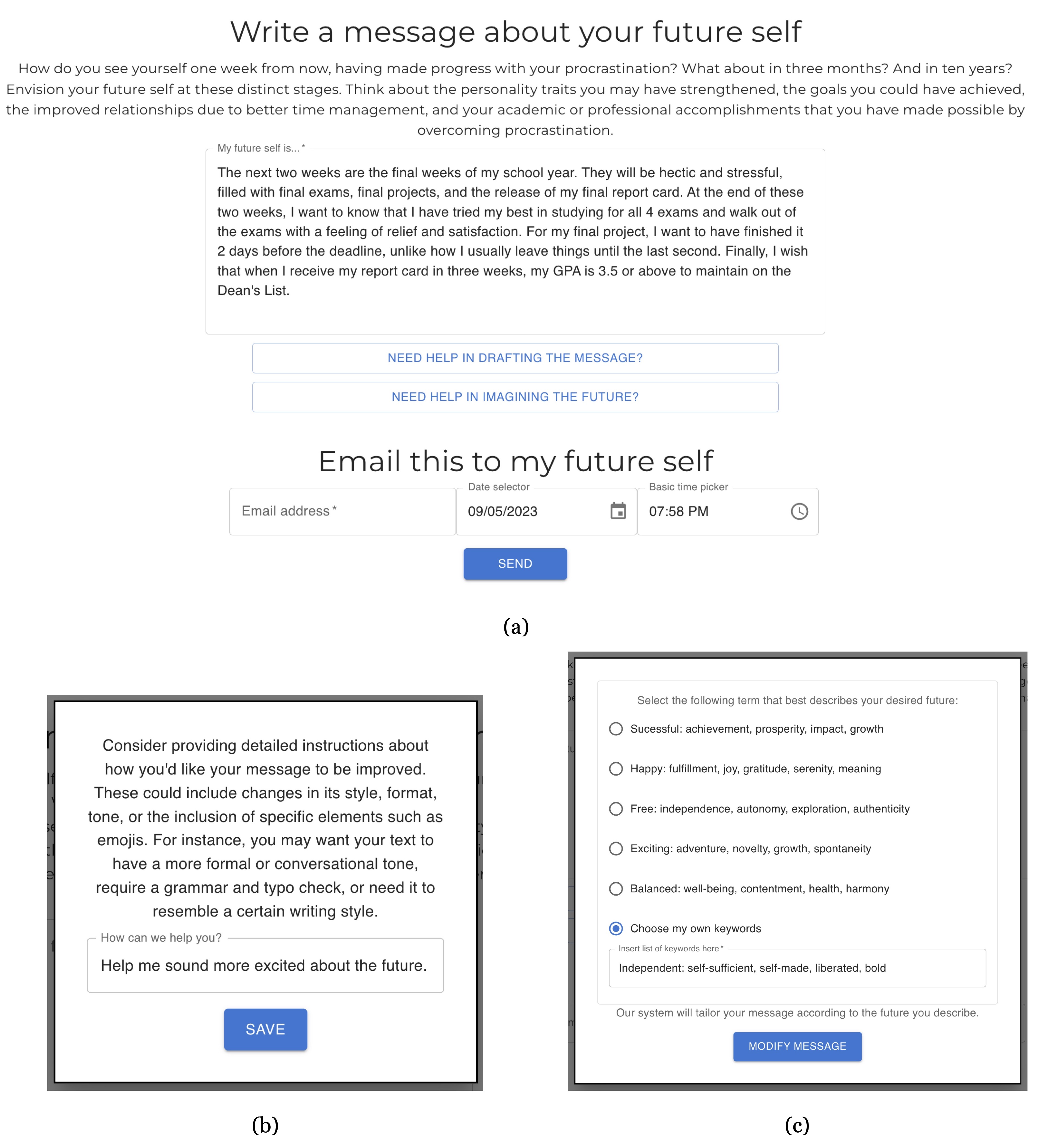}
\caption{Features for drafting an email to future self: (a) an open-ended text box for drafting the message, (b) a feature enabling open-ended modifications to the draft, (c) a list of suggested keywords for envisioning the future.}
\label{fig: future action}
\end{figure}

\subsection{Prompting Future Action} 
To foster proactive behavior and facilitate the translation of intentions into tangible actions, we implemented a feature that requested participants to draft an email to their future selves. Drawing on existing literature in the areas of task management and goal-setting \cite{grover2020design, agapie2022longitudinal, sheeran2016intention}, this feature aimed to encourage participants to prioritize their responsibilities and establish clear objectives. Our aim was to facilitate reflection at various temporal stages of their goals, from the immediate future to more distant timeframes, thereby enabling incremental action planning. In addition to aiding in the planning process, this feature also prompted participants to consider their underlying motivations and anticipated achievements \cite{pereira2021struggling}. 

While participants were given the autonomy to draft their future-oriented email independently, the LLM was available to assist in structuring and refining their thoughts. Participants had the flexibility to make open-ended requests for editing their drafts. They could also utilize specific keywords to conceptualize their future actions more effectively. These keywords were provided to serve as semantic anchors that can help individuals better align their actions with their core values and goals, thereby enhancing the potential for successful implementation \cite{gollwitzer2006implementation}.
These features are illustrated in Figure \ref{fig: future action}.


\subsection{Instructions and Examples}
\label{instructions}
Throughout the probe, participants were offered instructions and sample inputs for open-ended response boxes. The aim of providing these guidelines was to steer the participants toward relevant and effective suggestions \cite{herring2009getting, pearce1992case}. These examples and instructions were also formulated through multiple iterations with involvement from research team members with expertise in cognitive psychology. For instance, when participants were asked to describe a specific situation, as shown in Figure \ref{fig: buttons}(a), they were given example scenarios that they could follow for writing their own situation. Figures \ref{fig: buttons} and \ref{fig: future action} present the instructions and examples that were shown to the participants.

\begin{figure}[h!]
\centering
\includegraphics[width=8cm]{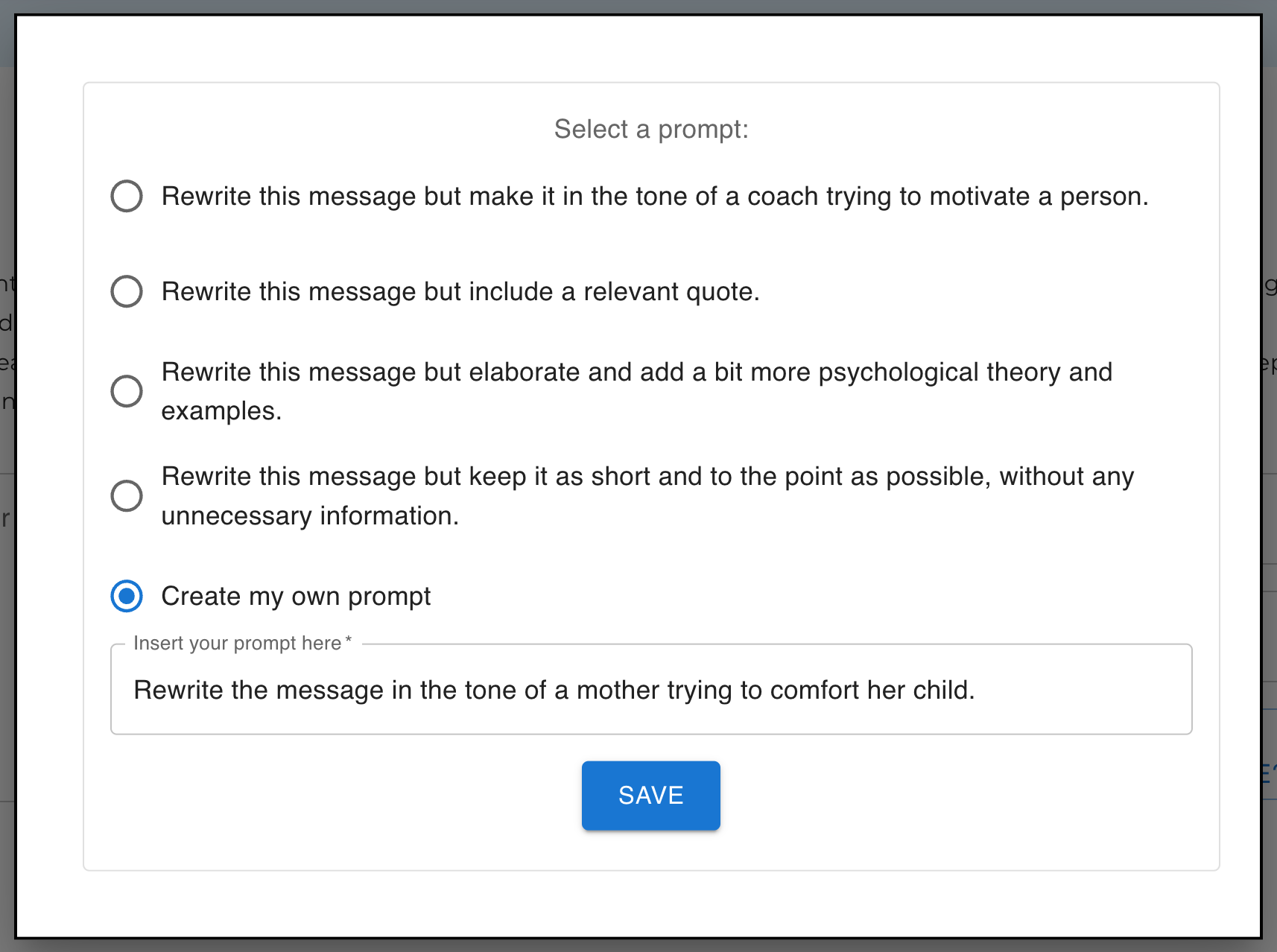}
\caption{Prompts to generate personalized messages}
\label{fig: prompt}
\end{figure}

Additionally, we recognized that some participants might desire greater autonomy in shaping the messages generated by the system, such as specifying the source of the advice or requiring the inclusion of particular elements \cite{bennett2023does, calvo2014autonomy}. To accommodate this, we enabled participants to create their own custom prompts. This feature allowed participants to explore how varying the input prompts can lead to diverse message outputs. To assist participants in crafting these prompts, we supplied a set of example templates that they could use as a starting point. These example prompts are displayed in Figure \ref{fig: prompt}.

\section{Methods}
We involved 15 university students in our study through interviews and focus group discussions and subsequently corroborated our findings through interviews with six experts in clinical psychology, learning and education, and cognitive psychology. The following section provides details on the methodology and logistics that shaped our study.

\subsection{Participants}


We recruited 15 individuals for our study through email invitations and social media calls. Participants were required to be between the ages of 18 and 25, currently enrolled at a university, and residing in North America. Additionally, they had to self-identify as at least average procrastinators, scoring 24 or higher according to the Irrational Procrastination Scale (IPS) \cite{svartdal2017measuring}. The mean age of our participants is 21.6$\pm$0.6 years old. They identified with two genders (10 women, 5 men; other options were offered) and several racial groups (8 Asian, 4 White, 2 African American, 1 Mixed-Race). We refer to these participants as P1--P15.

To obtain a professional perspective on participant feedback, we also conducted interviews with six experts. We refer to these experts as E1--E6. These participants were also recruited from email invitations and social media calls. All of them had post-graduate degrees from diverse fields including clinical psychology (E1 and E2), learning and education (E3--E6), and cognitive psychology (E4 and E6). They identified with two genders (3 women and 3 men; other options were offered) and belonged to two racial groups (5 White and 1 Asian).

\subsection{Procedure}

Our research protocol was approved by the Research Ethics Board at the first author's institution.
Recognizing the sensitive nature of our research topic, we anticipated that some participants might be more comfortable sharing their experiences and preferences in a one-on-one setting, while others might find a group discussion more conducive for collaborative idea generation. Therefore, we collected feedback through both individual interviews and focus group discussions. Ten individuals gave individual interviews (P1--P10) and five took part in a focus group discussion session (P11--P15). Two members of the research team organized the interviews and focus groups using the Google Meet videoconferencing platform. Interview durations ranged from 30 to 50 minutes, while the focus group discussion lasted 90 minutes. 

\revision{The interviews and focus groups employed a semi-structured approach to facilitate flexible investigation. After being given access to SPARK through a publicly hosted link, participants were given a brief overview of the study's objectives and a walkthrough of the tool according to the four main components as detailed in Section~\ref{ref: design}. Participants were then instructed to share their screens as they explored SPARK on their own for 10 minutes; some participants requested and were granted additional time for this part of the protocol (P1, P3, P4, P6, P9, and P10). While this exploration was self-directed, the typical user interaction proceeded as follows:
\begin{enumerate}
    \item Participants initially experimented with various versions of the seed messages (Figure \ref{fig: seed message}). This initial exploration typically lasted between 30 seconds and 2 minutes.
    \item They then explored different interface options as shown in Figures \ref{fig: buttons}(a) and \ref{fig: prompt}. Participants often sought clarification on the extent of detail required for their inputs, to which we clarified that they were free to provide as little or as much as they wanted. This phase of interaction took 6 to 8 minutes.
    \item The next step involved participants generating personalized messages using SPARK. They either requested further customization from the LLM or modified the messages themselves, as depicted in Figures \ref{fig: buttons}(b) and (c) respectively. This step usually spanned 2 to 4 minutes.
    \item Finally, participants engaged with the email drafting activity for future actions (Figure \ref{fig: future action}). Six participants composed short emails to be sent within the session's duration, while others experimented with different features. Participants spent  2 to 4 minutes on this activity.
\end{enumerate}
It is important to note that participants sometimes navigated non-linearly between the features of SPARK. Therefore, the times reported here represent the approximate duration for each stage of interaction.}


Next, participants were invited to give their thoughts on the utility of the tool and potential features they would find beneficial. The focus group session adhered to a similar structure, with the primary difference being that participants in the focus group were given turns to share their opinions. Questions posed in interviews and the focus group included, but were not limited to:

\begin{itemize}
\item What kinds of interface options or customization features would you like to see in a tool for managing academic procrastination?
\item How much effort are you willing to invest in detailing your procrastination issues in an open-ended text box?
\item What kinds of resources or guidance might help users effectively personalize messages using LLMs for managing academic procrastination?
\item How do you see such tools fitting into your daily routine for managing academic procrastination?
\end{itemize}

\noindent
All participants were compensated with \$20 USD per hour for their involvement in the study.

Following the completion of participant interviews, we applied thematic analysis to extract themes from the interview and focus group transcripts (Section \ref{sec:methods-analysis}). We then interviewed the experts to solicit feedback on these findings. These discussions also began with an introduction to the goals of our study and a walkthrough of SPARK, after which we presented a summary of the findings from our thematic analysis. For each theme, our questions for the experts centered around two broad questions:

\begin{itemize}
\item What are the potential benefits and risks of implementing the features related to this theme?
\item What ethical challenges or considerations should be anticipated when implementing these features?
\end{itemize}

The expert interviews were conducted via Google Meet and had durations ranging from 45 to 60 minutes. 


\subsection{Data Analysis}
\label{sec:methods-analysis}

\revision{The interviews and focus group discussions yielded approximately 9.5 hours of audio recordings, which translated into 172 pages of transcribed data. All personal identifiers were removed from the transcripts to ensure anonymity, and the data was then thoroughly cleaned for analysis. We followed a thematic analysis approach \cite{braun2006using, cooper2012apa, mcdonald2019reliability} to analyze the collected data. Two team members took on the role of coders and reviewed all data to familiarize themselves. The coders then began the process of assigning specific codes to three interview transcripts using an open-coding approach \cite{khandkar2009open}. Each coder individually created an initial codebook.}  

\revision{We subsequently employed a consensus coding approach akin to the process described by \citet{braun2006using}. The coders engaged in multiple collaborative sessions where they compared and merged their individual codebooks. This process of consensus coding with iterative discussions and refinement aimed to enhance consistency and reduce individual biases \cite{hill1997guide, mcdonald2019reliability}. Coders identified overlapping codes, clarified code definitions, and excluded codes not closely aligned with the research questions. The refined codebook was then applied iteratively to three additional transcripts, allowing for further adjustments based on emerging insights.
Once a final version of the codebook\footnote{\revision{The definitions of each code from the final codebook are provided in Appendix~\ref{app:code_def}.}} was agreed upon, each coder independently applied the established codes to half of the remaining data.}

\revision{Our analysis generated 19 codes, such as ``need for direct solutions'', ``step-by-step guidelines'', ``use of calendars'', and ``clear disclaimers''. The coders arranged multiple meetings to group the codes into broader themes through axial coding \cite{williams2019art}. We organized our findings around the themes that are presented in Section~\ref{sec: findings}.}

\subsection{Ethical Considerations}
Given the nature of our research, which touches upon aspects of procrastination, stress management, and goal setting, we were mindful of the ethical considerations involved. Throughout our work, we took measures to address these concerns.
To ensure participant comfort and safety, we informed them at the outset of the interviews that they had the option to skip any questions or discontinue the conversation if they felt uncomfortable at any point. Our interviewers were trained in the Columbia-Suicide Risk Assessment protocol \cite{posner2008columbia} and prepared to respond appropriately if participants expressed thoughts of suicide or self-harm. They were equipped to provide safety planning or make referrals to crisis services as necessary. However, none of these risks emerged during the study.
 
\section{Findings}
\label{sec: findings}
In the following section, we delineate various themes that emerge concerning the ways in which users perceive the utility of LLM-based tools for managing academic procrastination. Additionally, we discuss how expert opinions either corroborate or diverge from these user perspectives. \revision{Table~\ref{tab:thematic_analysis} shows the correspondence between themes and codes as well as the number of participants and experts who mentioned them.}

\begin{table}[]
\caption{\revision{Breakdown of themes from participant (P) and expert (E) responses.}}
\label{tab:thematic_analysis}
\centering
\begin{tabular}{|P{0.25\linewidth}|p{5cm}|p{3.5cm}|}
\hline
\textbf{\revision{Themes}} & \textbf{\revision{Codes}} & \textbf{\revision{Number of Individuals Mentioned}}\\\hline
\revision{Aspirations for structured action steps}  & \revision{Need for direct solutions} & \revision{13P + 2E} \\\cline{2-3}
 & \revision{Step-by-step guidelines} & \revision{10P + 4E} \\\cline{2-3}
 & \revision{Structuring recommendations} & \revision{10P + 3E} \\\cline{2-3}
 & \revision{Overreliance} & \revision{2P + 5E} \\\cline{2-3}
 & \revision{Adaptive breakdown} & \revision{3P + 2E} \\\hline
\revision{Deadline-driven instructions} & \revision{Reliance on deadlines} & \revision{12P + 4E} \\\cline{2-3}
 & \revision{Use of calendars} & \revision{7P + 2E} \\\cline{2-3}
 & \revision{Documenting goals} & \revision{8P + 5E} \\\cline{2-3}
 & \revision{Need for reminders} & \revision{7P + 4E} \\\cline{2-3}
 & \revision{Diversity in motivation} & \revision{2P + 4E} \\\hline
\revision{Guided versus unguided questions} & \revision{Varied levels of guidance} & \revision{6P + 5E} \\\cline{2-3}
 & \revision{Providing multiple layers of features} & \revision{7P + 3E} \\\cline{2-3}
 & \revision{Flexible engagement} & \revision{3P + 2E} \\\hline
\revision{Concerns and boundaries in} & \revision{Need for emotional support} & \revision{9P} \\\cline{2-3}
\revision{providing emotional support}  & \revision{Limitations of LLMs in addressing emotions} & \revision{1P + 6E} \\\cline{2-3}
 & \revision{Clear disclaimers} & \revision{5E} \\\hline
\revision{Providing support on the use of LLM-based tools} & \revision{Examples and instructions} & \revision{5P + 4E} \\\cline{2-3}
 & \revision{Collaborative elements} & \revision{4P + 2E} \\\cline{2-3}
 & \revision{Feedback to LLMs} & \revision{5P + 1E} \\
\hline
\end{tabular}
\end{table}

\subsection{Aspirations for Structured Action Steps}
Participants generally appreciated the flexibility offered by the tool (e.g., the ability to select a preferred tone or directedness) for message customization. 
They noted that customization not only enhanced the utility of the tool but also gave them a greater sense of control over their interaction. P1 commented:

\italquote{I think the tone and the length I’d love to play around with, I’m trying to imagine myself as a user. ChatGPT has a tendency to maintain the same tone in all of its messages. I don't think I'm going to the effort of putting this information there. But here people can just directly choose what they want to click in to use.}

Participants also raised potential ideas for additional customization features. For example, some vouched for the ability to switch between action-oriented and suggestive tones, and others suggested the ability to enhance instructions with pertinent resources (e.g., links to studies or relevant websites).  P12 stated that sometimes users may express a desire to understand the reasoning behind a step or to explore alternative approaches, so making such resources available could satiate this curiosity and augment user involvement. However, despite these design opportunities identified by participants, some of them also expressed the concern that an excessive number of options might be overwhelming. To address this, P2 suggested a two-tiered approach: presenting basic customization features on the main page and more advanced options on a separate page.

Extending this need for customization, participants in our study wished to receive structured guidance from the system to carry out their tasks. Many of them emphasized the challenges they face with unstructured or expansive tasks. The widespread view was that tasks appear more manageable when presented in a clear, step-by-step sequence. 
P8 drew upon a cooking analogy, referencing their preference for following a recipe versus freestyle cooking without any guidance. Beyond ease of use, participants also identified psychological benefits to seeing a concrete action plan. For instance, P2 noted how users could feel fulfillment upon completing individual steps for larger tasks.

Expert insights also underscored the value of offering users assistance in breaking down larger tasks. E1, however, cautioned that while a comprehensive plan can be constructive, there is a risk of overwhelming students. Their recommendation was to detail the imminent steps while providing a broader outline for later tasks. As users navigate through the initial steps, future steps and details can be introduced.
E3 and E6 also raised concerns about fostering over-reliance on the tool. E3 expressed:

\italquote{While structured guidance is valuable, we must ensure students retain their ability for independent thought and action. The danger lies in spoon-feeding them precise answers or solutions. It's crucial for students to develop their own thinking and information-gathering skills. Relying too heavily on a step-by-step model, especially one presented by an LLM, might foster a false sense of accomplishment. Students might follow every instruction and still wonder why their outcomes don't match their expectations.}

In summary, while structured guidance was seen as vital for task management, experts recommended a balance should be struck to ensure learners' independence and critical thinking remain intact.

\subsection{Deadline-Driven Interactions}

Participants appreciated the tool's ability to allow them to draft future emails to themselves. This feature was seen as especially beneficial for managing important deadlines. For instance, P10 indicated they would leverage this capability as a way to document their goals, with the periodic emails serving as reminders to review and refocus on these objectives. P2 and P4 further elaborated that the act of drafting these emails would also assist them in gaining better clarity regarding their goals and plans.

For better management of deadlines, however, several participants expressed an interest in integrating a tool like ours into their daily workflow for on-the-spot, personalized advice and actionable plans. Participants like P2 and P3 suggested integrations with platforms they already use for professional purposes, such as Google Calendar, Microsoft Teams, and Slack. In this context, a recurrent suggestion was the addition of a feature allowing participants to specify task deadlines. They believed that by doing so, the tool would be better positioned to deliver timely and pertinent recommendations. P2 shared an illustrative perspective:

\italquote{I have a friend who will be like, `Yeah, you have this coming up, you have that coming up. So you better do this first and do that first and prioritize this because this is more work than others.' So the AI can somewhat do something similar where you can be like, `You have this coming up. I think you should prioritize this.' It's not like it is telling me to do it. \dots It's just that, like a gentle reminder or something of that fashion.}

Some participants offered thoughts regarding the optimal frequency and timing of reminders from the tool. For instance, P6 emphasized the importance of receiving motivational cues significantly ahead of an impending deadline to prevent last-minute rushes. They went on to suggest incorporating features that allow users to set task priorities and provide estimates for completion times, tailoring the tool's reminders to their individual needs.

Experts largely agreed with the participants but emphasized the diverse motivational needs people require at different times. E2 suggested multiple strategies to this end: for those who feel regret reflecting on past procrastination, a message like ``The past is in the past, focus on today''; for specific barriers, a response inspired by WOOP (wish, outcome, obstacle, plan) goal-setting model \cite{oettingen2015rethinking}; and for a lack of perceived purpose, affirmations connecting tasks to future goals or values. Furthermore, E1 posited that presenting an array of motivational prompts could enable users to select the one that resonates most with them.


\subsection{Guided Versus Unguided Questions}
In our prototype, participants had the flexibility to elaborate on their situations to whatever extent they desired. Participants embraced this feature as an opportunity for them to mold canned messages to fit their unique context.
However, opinions varied when discussing the nature and quantity of questions the tool used to probe users for more information. Some participants advocated for a guided approach featuring a sequence of open-ended questions, believing it would help frame their responses more effectively. P7 illustrated this preference by stating, 

\italquote{The first problem is `What is something you are putting off for today?' And then you would write, `I'm putting off x and y'. And then after you type and press [the continue button], it might prompt you to say, `How has that made you feel?' And then it might prompt you to say, `What made me feel this way?'.}

Conversely, other participants were concerned that a sequence of questions could serve as a new avenue for procrastination. P2 emphasized the potential to overwhelm users with numerous prompts, especially during initial interactions with the platform. Additionally, building on past experiences with self-management tools, Participant P4 noted that a barrage of questions in the midst of a busy schedule could also heighten stress levels. In scenarios like these, they predicted that users would prefer to receive personalized guidance that is both immediate and easy to access, without the need to navigate through a multitude of questions.

To mitigate these concerns, E1, E3, and E6 suggested the inclusion of adaptive contextual factors to adjust both the quantity and type of questions posed to the user. They recommended employing contextual variables like the time of day, day of the week, and the user's schedule to fine-tune how the LLM elicits more information about the user's situation. While these experts generally supported providing longer interactions informed by psychological literature \cite{schouwenburg2004perspectives, pychyl2013solving}, they also recognized that users should have the option for shorter interactions when experiencing higher levels of activity or stress.



\subsection{Concerns and Boundaries in Providing Emotional Support}

The recognition and acknowledgment of user emotions emerged as a significant theme, with diverse opinions shared by both participants and experts. They recognized that discussions of procrastination could often lead to revelations about mental states, including symptoms of anxiety and depression. In such cases, some participants expressed the need for emotional support. P4 said:

\italquote{Assuming I already signed up for it, then I might like more emotional support. Like actual links [saying] `Here are resources that you can use'. Maybe there will be options [for showing] pep talks or something like that.}

\noindent
Participants like P7 felt that the tool could be proactive in inquiring users about their emotions and the possible culprits behind negative feelings. P11 went further, expressing an interest in having access to resources for managing more serious and vulnerable issues such as depression.

Nevertheless, experts advised caution regarding the prospect of LLMs taking on roles akin to therapists for providing support in the management of emotions. Their concern stemmed from the possibility that these models could dispense advice that, while perhaps appealing to users, may lack clinical validity. Instead, experts like E1, E4, and E5 emphasized that the tools should only simply recognize users' emotions while clearly communicating their limitations in providing therapeutic support. E3 said:

\italquote{First, acknowledge that the individual is struggling. You know, thank them for the description kind of thing. \ldots Just say this, `This kind of falls out of the scope of what I am able to assist you with'. }

\noindent
E5 elaborated on the need for transparency regarding the tool's limitations, emphasizing that users should be clearly informed from the outset about the tool's capabilities and constraints. This includes an explicit declaration at the introduction page outlining the tool's capacities and the message that any unintentional suggestion regarding emotion management should not be acted upon by users. They further stated that the system should contain contact information for university support services or crisis management centers for further assistance and support.

\subsection{Providing Support on the Use of LLM-based Tools}
Although LLMs are meant to be conversational and intuitive, participants in our study still desired guidance on how to best phrase their responses to get the tool to behave as they anticipated. Some felt that video tutorials and demonstrations might be useful, but there were stronger calls for concrete examples of inputs and outputs. P6 noted that the example prompts in Figure \ref{fig: prompt} helped them learn about the kinds of requests they could make and showed the range of possible outputs they could expect.
Extending on this idea further, participants proposed the idea of adding a collaborative element so that users could help one another optimally use the tool. P2 expanded on this concept by saying:

\italquote{Maybe share with me what other people are saying as well. So yeah, make it more like a social thing. I would also want to see what other people are asking for in their messages.}

Participants also expressed an interest in providing feedback on the advice generated by the tool so that it would work better for them in the future. Again, they were unsure about how this could be accomplished, so they sought instructions and concrete examples that could assist them in this aspect as well. P2 and P9 went on to suggest that such tools should offer users ways of providing feedback beyond inputs to the LLM. For instance, they envisioned a star-rating system that allows users to evaluate tailored advice on perceived qualities such as relevance, applicability, and level of detail.

Experts concurred with the participants' suggestions, adding an extra layer of caution. E5 stressed that tools need to explicitly distinguish the contexts in which they can be helpful from those in which they cannot. Accordingly, the examples provided to users should be designed to make this distinction abundantly clear so that users are fully aware of the tool's capabilities and limitations.
\section{Discussion}


\revision{In our discussion, we reflect on how our findings answer our research questions and then use those insights to generate design recommendations for LLM-based tools for managing academic procrastination. We then highlight our works' contributions to the broader literature on human-LLM collaboration. Finally, we enumerate the limitations of our study along with avenues for future research.}


\subsection{\revision{RQ1: Users' Envisioning of LLMs in Tailoring Strategies for Managing Academic Procrastination}}


\subsubsection{\revision{Structured Guidance with Adaptability}}
\revision{In alignment with existing literature on procrastination management and structured goal-setting \cite{pereira2021struggling, grover2020design, brechmann2013remind, das2023focused, hsieh2008using, kovacs2019conservation}, our study reiterates the importance of breaking down larger tasks into more manageable components. LLMs emerge as particularly adept for this purpose, offering the capability to provide detailed, actionable steps for immediate tasks while also delivering overarching, abstract guidelines for long-term objectives \cite{lin2022applying, mogavi2023exploring}. This dual approach of LLMs caters to both immediate task execution and longer-term goal planning.} 

\noindent~\\
\revision{\textbf{Design Recommendations:} Future procrastination management tools should strive to achieve a balance between providing structured guidance and maintaining adaptability. Ensuring that tools are both systematic in their approach to task management and flexible enough to adapt to individual user requirements could support users in a wide range of contexts \cite{bhattacharjee2023investigating, liffiton2023codehelp, kazemitabaar2023studying, denny2023promptly, leinonen2023using}.
For LLM-based tools specifically, an important feature could be the development of a \textit{dynamic structured guidance system}, which would generate hierarchical task lists dynamically based on user input \cite{o2018suddenly, lin2022applying, mogavi2023exploring}. Including functionality to automatically break down large tasks into manageable micro-tasks with clear deadlines would further enhance this system. Such a feature would not only provide structured task management but also adapt to the specific needs of each user. To enhance user engagement and provide comprehensive support, the system could also offer supplemental resources that are tailored to each specific task \cite{sallis2015ecological}.}

\subsubsection{\revision{Integration with User Routines and Deadlines}}
\revision{Participants in our study highlighted the necessity for deadline-oriented interactions facilitated by LLMs as a means
of seamlessly integrating into their daily academic activities. 
However, it is important to acknowledge the individual variations in perceived effort related to both the target task and other concurrent responsibilities. This aspect of our findings enhances and broadens the scope of existing literature, which underscores that effective deadline management should encompass more than just scheduling; it needs to take into account the comprehensive context of a user’s life \cite{edwards2015examining, martin2015effects, baker2016randomized, humphrey2021exploring, ye2022behavioral, zavaleta2022can}. Building on this, our research advocates for a dynamic approach to task prioritization, one that is attuned to the urgency of deadlines while being sensitive to the individual’s overall workload and personal commitments.}


\noindent~\\
\revision{\textbf{Design Recommendations:} Future procrastination management tools could benefit from synchronization with commonly used platforms such as Google Calendar, Microsoft Teams, and Slack. This integration would facilitate the automatic import of scheduled tasks and deadlines, thereby aiding in task organization \cite{howe2022design, grover2020design, das2023focused}. Additionally, integrating mobile phone sensor data could provide further information about users' activity levels and sleep patterns, potentially enhancing the relevance of task management suggestions \cite{ravichandran2017making, daskalova2020sleepbandits}. Based on this imported information and any user-inputted priorities, LLM-based tools could then dynamically generate a ranked task list. These tools could also include a \textit{effort estimator} to allow users to adjust task rankings based on their estimated completion time and task difficulty~\cite{agapie2022longitudinal}. }


\subsubsection{\revision{Collaborative Support in Using LLM-based Tools}}
\revision{Participants of our study expressed interest in learning from others' examples to both navigate our LLM-based tool and to learn about effective strategies for combating procrastination. 
This reflects the broader psychological need for collaborative support in combating procrastination \cite{uzun2013reducing, tuasikal2019role, hsueh2020exploring, burgess2019think}.
Implementing such collaborative support features could aid users in navigating the complexities of interacting with the LLM-based tool and offer them guidelines or best practices through the experiences of others \cite{bhattacharjee2022kind}. }

\noindent~\\
\revision{\textbf{Design Recommendations:} Future procrastination management tools could provide users with a communal repository where they could access a variety of strategies and solutions that have been effective in managing tasks and deadlines, bearing in mind that user contributions would need to be thoroughly de-identified to respect people's privacy \cite{thieme2023foundation}. 
Specifically for LLM-based tools, this concept could be further developed into a shared prompt library where users could view and contribute diverse examples of prompts and the corresponding responses generated by the LLM. This feature would be particularly useful as it would allow users to explore the range of the tool's capabilities, providing insights into how to interact more effectively with it. Furthermore, allowing users to rate the utility, relevance, and clarity of the LLM's responses in the communal repository could provide an effective means for users to identify the examples that best suit their needs.}



\subsection{\revision{RQ2: Potential Limitations and Tensions}}

\subsubsection{\revision{Boundaries in Addressing Emotions}}
\revision{Our study sheds light on the limitations of LLMs in the realm of procrastination management, particularly concerning emotional and therapeutic support. These findings align with recent skepticism among clinicians regarding the suitability of LLMs for emotional counseling and therapeutic support \cite{kasneci2023chatgpt, jo2023understanding, mogavi2023exploring}. While LLMs inherently lack the ability to discern human emotions, they might encounter user-expressed emotions in written form. The experts consulted in our study underscored the need for LLM-based tools to acknowledge and validate these emotions  without crossing into the territory of clinical or therapeutic advice \cite{jo2023understanding}. Their comments reinforce the importance of carefully defining the scope and limitations of LLM-based tools in managing tasks and emotions, ensuring their responsible and effective use.}

\noindent~\\
\revision{\textbf{Design Recommendations:} Future procrastination management tools should prominently feature a detailed disclaimer on the landing page to ensure clarity and responsible usage \cite{bhattacharjee2023investigating, bhattacharjee2022kind}. This disclaimer should concisely state the purpose and limitations of the tool, particularly in relation to emotional and psychological support. 
In the context of LLM-based tools, disclaimers should highlight that while they can assist with procrastination management strategies like scheduling or time management, they are not equipped to provide therapeutic  support \cite{korngiebel2021considering, ghassemi2021false}. For users who might require specialized emotional or psychological assistance, the system should provide a place where users can access a vetted list of well-established resources for mental health support: informational websites, helpline numbers, or links to accredited mental health service providers \cite{bhattacharjee2023investigating, bhattacharjee2022kind}.}

\subsubsection{\revision{Balancing Guidance and Independent Exploration}}
\revision{The expert interviewees expressed concerns about the potential for students to become overreliant on LLM-based tools \cite{kasneci2023chatgpt, jakesch2023co, mogavi2023exploring}. They warned that easy access to a tool that breaks down complex tasks might inhibit the development of independent problem-solving skills. 
This aligns with recent literature that emphasizes the need for a balance between providing guided assistance and fostering independent problem-solving skills \cite{thieme2023foundation, abd2023large}.}

\noindent~\\
\textbf{\revision{Design Recommendations:}} 
\revision{
Future procrastination management tools may occasionally restrict users' access to functionalities that produce direct step-by-step solutions and instead prompt them to independently seek out potential strategies \cite{cirillo2018pomodoro, kim2016timeaware}. These restrictions could be strategically timed, for instance, after a user completes a certain number of sub-tasks or reaches key milestones in a project. LLM-based tools could go a step further by generating \textit{independent exploration prompts} (e.g., ``research further'', ``think critically'' ) that encourage problem-solving strategies \cite{loksa2016programming, jin2019solvedeep}. Each option could come with a brief description of what the activity entails and why it might be beneficial, providing a nudge without forcing a specific course of action \cite{bhattacharjee2023design}. 
Users could also have the option to schedule independent exploration prompts at intervals of their choosing, offering customization to better suit individual work habits and needs \cite{bhattacharjee2023design}.}

\subsubsection{\revision{Allowing Flexible Engagement}}

\revision{We recognize a design tension regarding the extent of guidance users may wish to see when describing their situation. In some instances, users may prefer a set of structured questions to shape their responses, which may be more feasible during idle periods and moments of downtime \cite{poole2013hci, bhattacharjee2023investigating}. Conversely, a single open-ended question could be more suitable when users are initially acquainting themselves with the system or are in the midst of many other activities \cite{kornfield2020energy}. The same contextual variability applies to the level of customization available in the tool, whether it be related to the tone, directedness, or length of the advice given \cite{malle2001attention, bandura1994self, gnambs2010color}. While these features may be beneficial, there are moments when the user may not have the bandwidth to engage with such granularity and would prefer the option to bypass these decisions to avoid overwhelm \cite{bhattacharjee2023investigating}. }

\noindent~\\
\revision{\textbf{Design Recommendations:} Our findings highlight the need for procrastination management tools that permit flexible engagement, allowing users to maintain a minimal level of interaction even when they are not fully engaged. Studies in educational psychology and social media behaviors indicate that different online activities demand various levels of cognitive involvement \cite{bhattacharjee2023investigating, burke2010social, verduyn2021impact, haidet2004controlled}. While active engagement is typically more beneficial for enhancing learning and wellbeing, our findings suggest that there are times when users may be reluctant to invest significant cognitive effort, particularly during periods of low energy or mood \cite{bhattacharjee2023investigating}. In such instances, LLM-based tools should allow users to access key functionalities without demanding intensive engagement. 
While our findings are situated within the use of an LLM-based probe, this need for flexible engagement could apply to general procrastination management tools as well.}

\subsection{\revision{Contributions to Human-LLM Collaboration}}

\revision{Our findings, although mainly situated within the context of managing academic procrastination, contribute to the evolving body of literature on human-LLM collaboration. This study highlights the nuanced ways in which users engage with LLM-based tools, underscoring the potential for these models to offer more than just information processing; they can act as collaborative partners in personal and academic development \cite{chan2023mango, denny2023conversing, kim2022learning, guingrich2023chatbots, kumar2023math}. The desire for structured action steps and deadline-driven interactions aligns with the broader understanding of how individuals seek both assistance and partnership in managing their tasks. This perspective is bolstered by recent studies showing the promise of LLMs in exhibiting social intelligence and human-like collaborative behaviors \cite{park2023generative, shanahan2023role, zhao2023more, guingrich2023chatbots}. With this paradigm shift in AI that is moving towards systems that are capable of complex and empathetic interactions, there is a need to understand the unique contexts of individual users in order to enhance their downstream utility.} 

\revision{The varied opinions on structured questioning and the concerns raised about emotional support challenges resonate with ongoing discussions about the ethical and practical implications of human-LLM interactions \cite{thieme2023foundation, jo2023understanding, spatharioti2023comparing}. Our findings suggest that while users appreciate the personalized assistance offered by LLMs, they also value retaining a degree of autonomy and control over the interaction. This balance between guidance and independence could be a critical aspect of designing effective LLM-based tools, as it may directly impact user engagement and satisfaction \cite{bhattacharjee2023informing}. 
Moreover, the caution advised by experts about the role of LLMs in emotional support and the need for transparency in their capabilities reflects a growing awareness of the ethical responsibilities inherent in deploying AI in sensitive areas like mental health \cite{kasneci2023chatgpt, jo2023understanding, mogavi2023exploring}. The ethical implications of AI tools offering mental health support are complex, particularly given the risk of users relying on these tools for serious mental health issues without professional guidance \cite{kasneci2023chatgpt}. 
Our study contributes to the broader discourse on the ethical deployment of AI in sensitive domains, advocating for a cautious, well-informed approach that prioritizes user safety and wellbeing while exploring the beneficial aspects of human-LLM collaboration.
}


\subsection{Limitations and Future Works}

In acknowledging the limitations of our research, we first note that our participant group was composed of individuals residing in North America at the time of the study. Although this cohort included people from various racial groups, the geographical and cultural context in which they live may have influenced their perspectives on the use of LLM-based tools for managing procrastination. \revision{Future studies should aim to include a more geographically and culturally diverse participant group. Engaging with populations from different regions and cultural backgrounds would provide a broader perspective on the use of LLM-based tools for managing procrastination, potentially uncovering unique challenges, preferences, and expectations.}

Additionally, we designed a technology probe to explore user expectations regarding LLM-based tools for managing academic procrastination. To mitigate academic and emotional risks that could arise from deploying an unvalidated tool, we chose not to conduct a longitudinal study without researcher oversight. Nevertheless, we recognize the potential benefits of long-term studies for a more comprehensive understanding of the real-world effectiveness and long-term implications of LLM-based tools in this context.
\revision{With appropriate researcher oversight and ethical considerations, future studies could entail longitudinal deployments that examine the impact of various features on acceptability and perceived effectiveness.}

\revision{Finally, our findings are situated in the domain of academic procrastination management and, therefore, might not directly translate to other behavioral or psychological areas. Future research could broaden its scope to assess the efficacy of LLM-based tools in tackling a diverse array of challenges. Such exploration could range from addressing mental health concerns to enhancing personal productivity or promoting healthy behaviors. Investigating the application of these tools in varied scenarios will provide a deeper understanding of their adaptability and constraints, enriching the knowledge base about the full potential of LLM-based tools in different contexts.}
\section{Conclusion}

Traditional interventions often fail to capture the unique, person-specific factors that contribute to academic procrastination. LLMs offer a promising avenue to bridge this gap, as they allow for open-ended, individualized responses through sophisticated data analysis and natural language processing capabilities. Yet, the expectations and constraints of employing LLMs for this purpose have not been thoroughly investigated. To fill this knowledge void, we conducted in-depth interviews and focus group discussions with 15 university students and consulted six experts, using a technology probe for context. Our findings underscore the need for LLM-based tools to offer structured guidance with a focus on deadlines and to include enhanced mechanisms for user support. The results also emphasize the importance of adaptability in the query process, taking into account factors such as the individual's level of busyness, and advocate for caution when considering the use of LLMs for therapeutic advice. These insights enabled us to make several design recommendations surrounding structured guidance with adaptive complexity, boundaries in addressing user emotions, and collaborative support in using LLMs. We believe our findings and recommendations will serve as important guidelines for developers, UX/UI designers, and researchers working on LLM-based intervention tools.

\bibliographystyle{ACM-Reference-Format}
\bibliography{sample-base}
\appendix

\section{Code Definitions}
\label{app:code_def}

Our final codebook had 19 codes in total. We provide their definitions below:

\begin{itemize}
    \item \textbf{Need for direct solutions:} Preferences for straightforward, practical advice or solutions that directly address various aspects of academic procrastination.
    \item \textbf{Step-by-step guidelines:} The desire for detailed instructions that are laid out in a sequential order.
    \item \textbf{Structuring recommendations:} Well-organized, systematic advice that helps in planning and organizing tasks. Structuring recommendations could involve breaking down large tasks into smaller, manageable parts.
    \item \textbf{Overreliance:} Concerns about the possibility of becoming too dependent on Large Language Models for solutions, potentially diminishing one's own problem-solving skills.
    \item \textbf{Adaptive breakdown:} The desire for advice that can be customized or broken down according to the individual's specific needs and circumstances.
    \item \textbf{Reliance on deadlines:} Making recommendations around important deadlines.
    \item \textbf{Use of calendars:} Incorporating calendars as tools for organization, planning, and keeping track of deadlines and tasks.
    \item \textbf{Documenting goals:} The practice of writing down goals as a means to keep track of progress.
    \item \textbf{Need for reminders:} The desire for prompts or alerts to help individuals remember and adhere to their planned tasks and deadlines.
    \item \textbf{Diversity in motivation:} Presenting a variety of motivational prompts to users to help them maintain motivation over time.
    \item \textbf{Varied levels of guidance:} The need for different levels of guidance and support, based on the preferences of individuals and their circumstances.
    \item \textbf{Providing multiple layers of features:} Offering a range of features that allow for varied forms of interaction and support, catering to different user preferences and needs.
    \item \textbf{Flexible engagement:} Allowing users to choose their level of interaction and engagement with the tool, accommodating different styles of learning and task management.
    \item \textbf{Need for emotional support:} Expressing a desire for assistance that goes beyond task management to include emotional support.
    \item \textbf{Limitations of LLMs in addressing emotions:} Recognizing that while LLMs can offer task-related support, their ability to provide emotional support is limited.
    \item \textbf{Clear disclaimers:} The importance of having explicit disclaimers about the capabilities and limitations of LLMs, especially regarding emotional support and therapeutic advice.
    \item \textbf{Examples and instructions:} The request for practical examples and clear instructions on how to effectively use LLMs for managing academic procrastination.
    \item \textbf{Collaborative elements:} The interest in features that enable collaboration, either with the LLM or with other users.
    \item \textbf{Feedback to LLMs:} Ways to integrate user feedback with LLM-based tools for improving their effectiveness and relevance.
\end{itemize}

\end{document}